# GPU-based ultra fast IMRT plan optimization

**Chunhua Men[1], Xuejun Gu[1], Dongju Choi[2], Amitava Majumdar[2], Ziyi Zheng[3], Klaus Mueller[3], and Steve B. Jiang[1]**

[1]Department of Radiation Oncology, University of California San Diego, La Jolla, CA 92037, USA
[2]San Diego Supercomputer Center, University of California San Diego, La Jolla, CA 92093, USA
[3]Center for Visual Computing, Computer Science Department, Stony Brook University, NY 11794, USA

E-mail: sbjiang@ucsd.edu

The widespread adoption of on-board volumetric imaging in cancer radiotherapy has stimulated research efforts to develop online adaptive radiotherapy techniques to handle the inter-fraction variation of the patient's geometry. Such efforts face major technical challenges to perform treatment planning in real-time. To overcome this challenge, we are developing a supercomputing online re-planning environment (SCORE) at the University of California San Diego (UCSD). As part of the SCORE project, this paper presents our work on the implementation of an intensity modulated radiation therapy (IMRT) optimization algorithm on graphics processing units (GPUs). We adopt a penalty-based quadratic optimization model, which is solved by using a gradient projection method with Armijo's line search rule. Our optimization algorithm has been implemented in CUDA for parallel GPU computing as well as in C for serial CPU computing for comparison purpose. A prostate IMRT case with various beamlet and voxel sizes was used to evaluate our implementation. On an NVIDIA Tesla C1060 GPU card, we have achieved speedup factors of 20-40 without losing accuracy, compared to the results from an Intel Xeon 2.27 GHz CPU. For a specific 9-field prostate IMRT case with 5×5 mm$^2$ beamlet size and 2.5×2.5×2.5 mm$^3$ voxel size, our GPU implementation takes only *2.8 seconds* to generate an optimal IMRT plan. Our work has therefore solved a major problem in developing online re-planning technologies for adaptive radiotherapy.

## 1. Introduction

The goal of radiation oncology is to deliver a prescribed radiation dose to targets containing tumor and cancerous regions while sparing surrounding functional organs and normal tissues. In current clinical practice, an "optimal" treatment plan is designed based on patient geometry acquired before the first treatment fraction and used for the whole treatment course. This practice implicitly (and erroneously) assumes that the patient anatomy is static and ignores the facts that 1) the position and shape of the tumor and nearby organs may vary substantially from day and day, and 2) there might be significant dynamic change in tumor geometry in response to the radiotherapy. As a result, the quality of this pre-designed treatment plan may deteriorate during the course of treatment. To address this issue, various adaptive radiotherapy (ART) technologies have been developed, with the basic idea of imaging the patient's changed geometry during the treatment course and modifying the treatment plan accordingly (Yan *et al.*, 1997; Wu *et al.*, 2002; Birkner *et al.*, 2003; Wu *et al.*, 2004; Mohan *et al.*, 2005; de la Zerda *et al.*, 2007; Lu *et al.*, 2008; Wu *et al.*, 2008; Fu *et al.*, 2009; Godley *et al.*, 2009).

One way to implement ART is called online ART, where the treatment plan is re-designed in a real-time fashion based on the patient's updated geometry to ensure the optimality of the treatment and to ensure that the treatment objectives are met (Wu *et al.*, 2002; Mohan *et al.*, 2005; Wu *et al.*, 2008; Fu *et al.*, 2009). The major challenge of this technology is that the re-planning process has to be done in a few minutes since the patient is lying on the treatment couch and waiting for the treatment. At the University of California San Diego (UCSD), we are developing a supercomputing online re-planning environment (SCORE) for online ART. The system consists of three major components: (1) patient modeling based on deformable image registration, (2) dose calculation, and (3) plan optimization. The general aim of the SCORE project is to parallelize, implement, and test the algorithms of those three major components on supercomputing platforms that are affordable and easy to maintain for regular radiotherapy clinics. One such platform that we are evaluating is a workstation PC accelerated with graphics processing units (GPUs).

Intensity modulated radiation therapy (IMRT) plan optimization is a large-scale optimization problem. The current state of the art optimization models implemented on a single CPU still at least take several minutes to optimize a treatment plan (Romeijn *et al.*, 2003; Men *et al.*, 2007; Scherrer and Kuefer, 2008; Wu *et al.*, 2008). Parallel computing potentially has the ability to increase the computation efficiency. One way is to use computer clusters or traditional supercomputers. However, they are not readily available to most clinical users. In addition, communication among CPUs can be expensive, leading to marginal efficiency gains for the IMRT optimization problem. The advent of multi-core processor PCs implementing multiprocessing in a single physical package makes parallel computing affordable for most users. However, the level of parallelism is very limited --- in the best case, the speedup factor gained by the use of a multi-core processor is near the number of cores. GPUs, on the other hand, have up to hundreds of processing cores, can be effectively used for parallel computing, and are affordable to most users. While GPUs were initially used primarily for graphics applications, they have recently been increasingly used for general purpose scientific computing (Jung and O'Leary, 2007; Munekawa *et al.*, 2008; Xing *et al.*, 2008; Su *et al.*, 2009; Wallner, 2009). For cancer radiotherapy, researchers have also started to use GPUs to accelerate some heavy computational tasks, such as CBCT reconstruction, deformable image registration, and dose calculation (Sharp *et al.*, 2007; Xu and Mueller, 2007; Noe *et al.*, 2008; Samant *et al.*, 2008; Hissoiny *et al.*, 2009). However, based on our knowledge, no work has been done to speed up treatment plan optimization using GPUs.

In this paper, we report our implementation of an IMRT optimization algorithm using Compute Unified Device Architecture (CUDA) on GPUs. The performance of the CUDA implementation will be evaluated against a corresponding C implementation on a CPU.

**2. Methods and Materials**

*2.1 Fluence map (re-)optimization model*

In IMRT optimization, each beam is decomposed into a set of beamlets (denoted by *N*). The set of voxels that represents the patient's CT image is denoted by *V*. In addition, we denote the dose to voxel $j \in V$ by $z_j$ and the intensity of beamlet $i \in N$ by $x_i$. The voxel dose is calculated as $z_j = \sum_{i \in N} D_{ij} x_i$, where $D_{ij}$ is called *dose deposition coefficients* and represents the dose received by voxel $j \in V$ from beamlet $i \in N$ of unit intensity. Finally, our fluence map (re-)optimization model employs treatment plan evaluation criteria that are quadratic one-sided voxel-based penalties. If we denote the set of target voxels by $V_T$, then we can write the criteria as:

$$F_j^-(z_j) = (\max\{0, T_j - z_j\})^2 \quad j \in V_T$$
$$F_j^+(z_j) = (\max\{0, z_j - T_j\})^2 \quad j \in V \quad , \tag{1}$$

where $T_j$ for all $j \in V$ represents the dose distribution in the original treatment plan. In this work, we are not trying to generate "the best plan of the day"; instead, we are trying to reproduce the dose distribution from the original plan within a tolerance by re-optimizing the fluence maps to accommodate the changed patient's geometry. The advantage of this approach is that, if the new plan is the equivalent to the original plan within a pre-specified tolerance, it can be used to treat the patient without the physician's re-approval. Later in the paper we will discuss alternative re-optimization models.

If we let $F_j(z_j) = F_j^-(z_j) \cup F_j^+(z_j)$, then our model can be written as:

$$\min \sum_{j \in V} F_j(z_j) \tag{2}$$

Subject to

$$z_j = \sum_{i \in N} D_{ij} x_i \qquad j \in V$$

$$x_i \geq 0 \qquad i \in N.$$

*2.2 The optimization algorithm*

The objective function is convex quadratic in our model, thus the direction of steepest descent is that of the negative gradient. However, since the decision variables, which are the intensities of beamlets, should be nonnegative, moving along the steepest descent direction may lead to infeasible intensities. We therefore use the gradient projection method (see, *e.g.*, (Bazaraa *et al.*, 2006)) which projects the negative gradient in a way that improves the objective function while maintaining feasibility. For convenience, we denote:

$$F_j(z_j) = F_j(\sum_{i \in N} D_{ij} x_i) = G_j(\mathbf{x}). \tag{3}$$

Then our model can be rewritten as:

$$\min G(\mathbf{x})$$
Subject to $\tag{4}$
$$x_i \geq 0 \quad i \in N.$$

The gradient projection method is an iterative method and the new solution is given by $\mathbf{x}_{k+1} = P(\mathbf{x}_k - \lambda_k \nabla G(\mathbf{x}_k)^T)$, where $\nabla G(\mathbf{x}_k)$ denotes the gradient of $G(\mathbf{x})$ at $\mathbf{x}_k$, $\lambda_k$ denotes the step size at iteration $k$, and $P$ denotes the projection onto the feasible field. That is, given $\mathbf{y} \in R^n$, $P(\mathbf{y}) = \arg\min_{\mathbf{x} \geq 0}(\|\mathbf{y} - \mathbf{x}\|)$, where $\|\cdot\|$ is the Euclidean norm. Since the nonnegativity constraints are the only ones that need to be satisfied in the model, we can simply obtain $P(\mathbf{y}) = \max(\mathbf{0}, \mathbf{y})$.

To obtain $\lambda_k$, we use Armijo's rule (see, *e.g.*, (Bazaraa *et al.*, 2006)), which balances the sufficient degree of accuracy and the convergence of the overall algorithm. Armijo's rule is driven by two parameters, $0 < \varepsilon < 1$ and $\alpha > 1$. At the beginning of the line search in each iteration, a fixed step size $\lambda$ is given, and the step size is acceptable if the new objective function is less than a threshold $C_k$. Otherwise, the step size is sequentially decreased by letting $\lambda = \lambda/\alpha$. In our model, the descent direction of line search is the negative gradient, and by derivation, we obtain the threshold as follows:

$$C_k = G(\mathbf{x}_k) - \frac{\varepsilon}{\lambda}(\mathbf{x}_{k+1} - \mathbf{x}_k)(\mathbf{x}_{k+1} - \mathbf{x}_k)^T. \tag{5}$$

We denote the stop tolerance by $\delta$. Our algorithm can be summarized as follows:

***Initialization***

   Select initial solution $\mathbf{x}_1$ satisfying $\mathbf{x}_1 \geq \mathbf{0}$.
   Select parameters $\varepsilon$, $\alpha$ and $\lambda_0$. Let $k = 1$.

***Main Iterative Loop***

1. Let $d_k = -\nabla G(\mathbf{x}_k)$.

2. $\lambda = \lambda_0$; $\boldsymbol{x}_{k+1} = \max(\boldsymbol{0}, \boldsymbol{x}_k + \lambda \boldsymbol{d}_k)$.
3. If $G(\boldsymbol{x}_{k+1}) < C_k$, go to step 5; else go to step 4.
4. $\lambda = \lambda/\alpha$; $\boldsymbol{x}_{k+1} = \max(\boldsymbol{0}, \boldsymbol{x}_k + \lambda \boldsymbol{d}_k)$; go to step 3.
5. If $|G(\mathbf{x}_k) - G(\mathbf{x}_{k-1})|/G(\mathbf{x}_k) < \delta$, stop; else $k = k + 1$, go to step 1.

In this work, we specify the stopping criterion as $\delta = 10^{-5}$.

*2.3 GPU implementation*

GPUs are designed specifically for computationally intensive, highly data-parallel applications. In November 2006, NVIDIA introduced CUDA, a general-purpose parallel computing architecture with a new parallel programming model and instruction set architecture. In the CUDA environment, a GPU has to be used in conjunction with a CPU. The CPU serves as the *host* while the GPU is called the *device*. The main code runs on the host, invoking *kernels* that are executed in parallel on the device by using a large number of CUDA *threads* (NVIDIA, 2009).

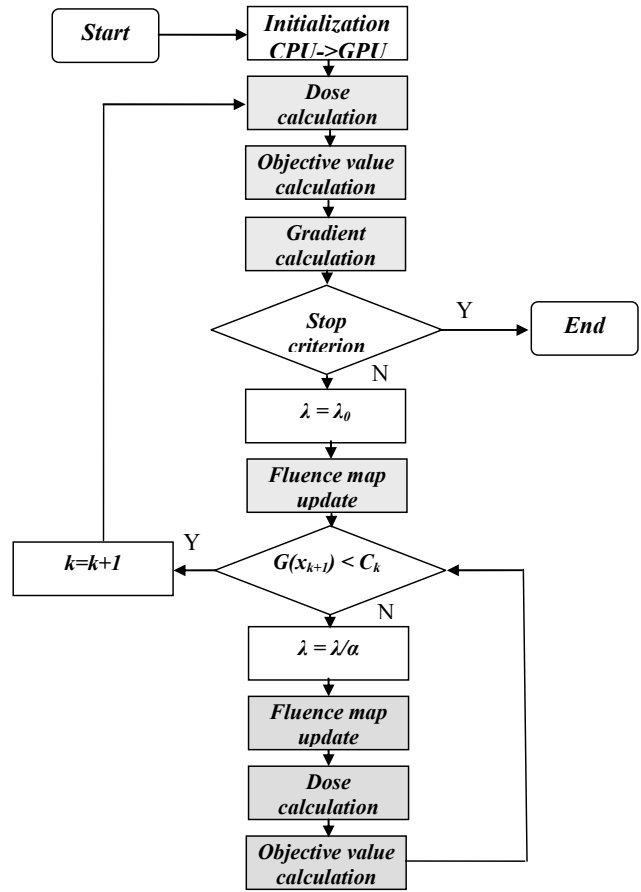

**Figure 1.** Flow chart of our GPU-based gradient projection algorithm. Grey boxes indicate CUDA kernels running on the device (GPU) and white boxes indicate C code running on the host (CPU).

Figure 1 depicts the flow chart of our GPU implementation of the gradient projection algorithm. The majority of the algorithm runs on the device as CUDA kernels (grey boxes), while the remaining simple arithmetic operations run on the host (white boxes). We use 4 kernels in our CUDA implementation: dose calculation, objective function calculation, gradient calculation, and fluence map update ($x_{k+1}$ = max($0$, $x_k + \lambda d_k$ )). For the dose calculation, each thread calculates the dose received by a certain voxel. For the objective function calculation, each thread calculates the objective value contributed by a certain voxel, which is summarized to a total objective value using parallel reduction and loop unrolling operations. For the gradient calculation, each thread calculates the gradient value at a certain bixel.

Some global variables, such as $x_k$, may be accessed by various threads simultaneously, causing massive memory accesses which can significantly impair the efficiency. To avoid this, we store those variables in texture memory which can be accessed as cached data.

The dose deposition coefficient matrix is a sparse matrix due to the fact that a beamlet only contributes to voxels close to its path. We therefore store the $D_{ij}$'s using the *compressed sparse row* (CSR) format, which is the most popular general-purpose sparse matrix representation. To improve the performance of our CUDA code for calculating the voxel dose, which is actually a sparse matrix-vector multiplication, we use a newly developed method (Bell and Garland, 2008) on GPU.

## 3. Experiments and Results

### 3.1 Clinical testing case

To test our implementation, we used one clinical case of prostate cancer with three scenarios of different beamlet/voxel sizes (Table 1). For each scenario, nine co-planar beams were evenly distributed around the patient and the prescription dose to PTV was 73.8 Gy. For target and organs at risk (OARs), we used a voxel size of 4×4×4 mm$^3$ in the first two scenarios and 2.5×2.5×2.5 mm$^3$ in the last scenario. For unspecified tissue (*i.e.*, tissues outside the target and OARs), we increased the voxel size in each dimension by a factor of 2 to reduce the optimization problem size. The full resolution was used when evaluating the treatment quality (does volume histograms (DVHs), dose color wash, isodose curves, etc.).

**Table 1.** Running times for plan optimization for CPU and GPU implementations tested on a clinical case with various beamlet and voxel sizes.

| # | # beams | beamlet size (mm$^2$) | # beamlets | voxel size (mm$^3$) | # voxels | # non-zero $D_{ij}$'s | CPU (s) | GPU (s) | Speedup |
|---|---|---|---|---|---|---|---|---|---|
| 1 | 9 | 10×10 | 2,055 | 4×4×4 | 35,988 | 3,137,805 | 3.81 | 0.19 | 20.1 |
| 2 | 9 | 5×5 | 6,453 | 4×4×4 | 35,988 | 10,612,611 | 16.4 | 0.49 | 33.5 |
| 3 | 9 | 5×5 | 6,453 | 2.5×2.5×2.5 | 143,329 | 43,266,357 | 111.8 | 2.79 | 40.1 |

*3.2 Test results*

We tested our CUDA implementation on an NVIDIA Tesla C1060 GPU card, which has 30 multiprocessors (each with 8 SIMD processing cores) and 4GB of memory. For comparison purposes, we also implemented our algorithm in sequential C code and tested it on an Intel Xeon 2.27GHz CPU. We used the same sparse matrix format to represent sparse matrix ***D*** for both implementations.

We first analyzed the accuracy of the results. The input data is single floating point precision on both CPU and GPU. The difference between the final objective function values calculated by CPU and GPU was around $10^{-6} \sim 10^{-5}$ which is negligible in clinical practice. In fact, by checking the DVHs, dose color wash, and isodose curves, no differences could be observed between CPU and GPU results. Figure 2 shows the DVHs and dose wash superimposed on a representative CT slice of this clinical case, both corresponding to an optimal treatment plan for Scenario 3 on GPU. The running times for plan optimization for the 3 scenarios on both CPU and GPU are shown in Table 1. The speedup achieved by the GPU implementation over the CPU implementation improved from 20.1 to 40.1 as the problem size increased from Scenario 1 to Scenario 3.

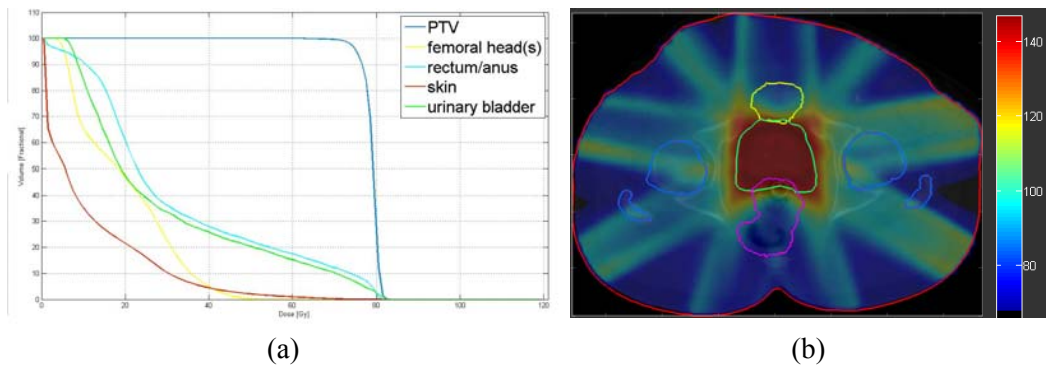

(a)                                                             (b)

**Figure 2**. The optimal treatment plan obtained for Scenario 3 on GPU: (a) DVHs; (b) dose color wash on a representative CT slice.

## 4. Discussion and Conclusions

This paper presents the implementation of an IMRT (re-)optimization algorithm on GPUs. The CUDA kernels exploit fine-grained parallelism to effectively utilize the computational resources of the GPU. We tested our implementation on a 9-field prostate IMRT case with various beamlet and voxel sizes, and our results show that an optimal plan can be obtained in 0.2 to 2.8 seconds using a Tesla C1060 card, and the speedup factor varies from 20 to 40 when using an Intel Xeon 2.27 GHz CPU as the reference platform.

We investigated the performance of our CUDA implementation on different GPUs with similar clock speeds, including NVIDIA's GeForce 9500 GT, GTX 285, Tesla

C1060, and Tesla S1070. The GeForce 9500 GT has only 4 multiprocessors and produced limited speedup (1x-3x). The GPUs on the GTX 285 and S1070 each have 30 multiprocessors (same as C1060) and delivered similar speedup results as described for the C1060 (Table 1). The memory required for the largest dataset tested (Scenario 3) is less than 1GB, which could be accommodated by all of the GPUs that were tested. We found, for these test scenarios, that the speedup factors solely dependent on the number of multiprocessors per GPU.

We also noticed that the speedup factor is different for different CUDA kernels. In Scenario 3, the speedup factor is 46 for the dose calculation kernel, 18 for the objective value calculation kernel, 12 for the gradient calculation kernel. The dose and objective calculations are parallelized based on voxels while the gradient calculation is parallelized based on bixels. In general, the number of voxels is much greater than the number of bixels. Therefore, it is not surprising to see that we can obtain better speed up for the dose and objective calculations. The speedup for the objective function value calculation is not as good as that for the dose calculation due to the communication cost of the parallel reduction algorithm.

In the re-optimization model presented in this paper, we are trying to reproduce the dose distribution from the original plan for each individual fraction. This model might be too rigid. We may not be able to reproduce the exact dose distribution for some new geometries. In that case, the accumulative dose distribution may differ significantly from the desired dose distribution. A better re-optimization model should take into account the accumulative dose distribution delivered in pervious fractions, by either using dose volume constraints or setting $T_j$ in Equation (1) as the difference between the desired accumulative dose distribution up to the current fraction and the delivered accumulative dose distribution in previous fractions. We will explore both approaches in our future work.

Multi-leaf collimator (MLC) leaf sequences have to be generated from our optimized fluence maps, which is usually very fast (less than 1 second on CPU). A major drawback of fluence map optimization followed by a leaf-sequencing stage is that there is a potential loss in the treatment quality. To overcome this, we plan to implement a direct aperture optimization (DAO) algorithm on GPUs (Men *et al.*, 2007).

**Acknowledgements**

We would like to thank NVIDIA for providing GPU cards and Hubert Pan for his constructive comments on the manuscript. This work is supported in part by the University of California Lab Fees Research Program. Ziyi Zheng and Klaus Mueller are supported by NSF grant CCF-0702699.

**References**

Bazaraa M S, Sherali H D and Shetty C M 2006 *Nonlinear Programming: Theory and Algorithms* (Hoboken: A John Wiley & Sons)

Bell N and Garland M 2008 Efficient sparse matrix-vector multiplication on CUDA. In: *NVIDIA Technical Report,* (Santa Clara: NVIDIA Corporation)

Birkner M, Yan D, Alber M, Liang J and Nusslin F 2003 Adapting inverse planning to patient and organ geometrical variation: algorithm and implementation *Medical Physics* **30** 2822-31

de la Zerda A, Armbruster B and Xing L 2007 Formulating adaptive radiation therapy (ART) treatment planning into a closed-loop control framework *Physics in Medicine and Biology* **52** 4137-53

Fu W H, Yang Y, Yue N J, Heron D E and Huq M S 2009 A cone beam CT-guided online plan modification technique to correct interfractional anatomic changes for prostate cancer IMRT treatment *Physics in Medicine and Biology* **54** 1691-703

Godley A, Ahunbay E, Peng C and Li X A 2009 Automated registration of large deformations for adaptive radiation therapy of prostate cancer *Medical Physics* **36** 1433-41

Hissoiny S, Ozell B and Despres P 2009 Fast convolution-superposition dose calculation on graphics hardware *Medical Physics* **36** 1998-2005

Jung J H and O'Leary D P 2007 Implementing an interior point method for linear programs on a CPU-GPU system *Electron. Trans. Numer. Anal.* **28** 174-89

Lu W G, Chen M, Chen Q, Ruchala K and Olivera G 2008 Adaptive fractionation therapy: I. Basic concept and strategy *Physics in Medicine and Biology* **53** 5495-511

Men C, Romeijn H E, Taskin Z C and Dempsey J F 2007 An exact approach to direct aperture optimization in IMRT treatment planning *Physics in Medicine and Biology* 7333-52

Mohan R, Zhang X D, Wang H, Kang Y X, Wang X C, Liu H, Ang K, Kuban D and Dong L 2005 Use of deformed intensity distributions for on-line modification of image-guided IMRT to account for interfractional anatomic changes *International Journal of Radiation Oncology Biology Physics* **61** 1258-66

Munekawa Y, Ino F and Hagihara K 2008 Design and implementation of the Smith-Waterman algorithm on the CUDA-compatible GPU *2008 8th IEEE International Conference on Bioinformatics and BioEngineering* 6 pp.

Noe K O, De Senneville B D, Elstrom U V, Tanderup K and Sorensen T S 2008 Acceleration and validation of optical flow based deformable registration for image-guided radiotherapy *Acta Oncologica* **47** 1286-93

NVIDIA 2009 NVIDIA CUDA Programming Guide 2.2

Romeijn H E, Ahuja R K, Dempsey J F, Kumar A and Li J G 2003 A novel linear programming approach to fluence map optimization for intensity modulated radiation therapy treatment planning *Physics in Medicine and Biology* **48** 3521-42

Samant S S, Xia J Y, Muyan-Ozcelilk P and Owens J D 2008 High performance computing for deformable image registration: Towards a new paradigm in adaptive radiotherapy *Medical Physics* **35** 3546-53

Scherrer A and Kuefer K H 2008 Accelerated IMRT plan optimization using the adaptive clustering method *Linear Algebra and Its Applications* **428** 1250-71

Sharp G C, Kandasamy N, Singh H and Folkert M 2007 GPU-based streaming architectures for fast cone-beam CT image reconstruction and demons deformable registration *Physics in Medicine and Biology* **52** 5771-83

Su C, Fu Z-l and Tan Y-c 2009 Fast operation of large-scale high-precision matrix based on GPU *Journal of Computer Applications* 1177-9, 92

Wallner G 2009 An extended GPU radiosity solver *Visual Computer* **25** 529-37

Wu C, Jeraj R, Lu W G and Mackie T R 2004 Fast treatment plan modification with an over-relaxed Cimmino algorithm *Medical Physics* **31** 191-200

Wu C, Jeraj R, Olivera G H and Mackie T R 2002 Re-optimization in adaptive radiotherapy *Physics in Medicine and Biology* **47** 3181-95

Wu Q J, Thongphiew D, Wang Z, Mathayomchan B, Chankong V, Yoo S, Lee W R and Yin F F 2008 On-line re-optimization of prostate IMRT plans for adaptive radiation therapy *Physics in Medicine and Biology* **53** 673-91

Xing M, Decaudin P, Baogang H and Xiaopeng Z 2008 Real-time marker level set on GPU *2008 International Conference on Cyberworlds* 209-16
Xu F and Mueller K 2007 Real-time 3D computed tomographic reconstruction using commodity graphics hardware *Physics in Medicine and Biology* **52** 3405-19
Yan D, Vicini F, Wong J and Martinez A 1997 Adaptive radiation therapy *Physics in Medicine and Biology* **42** 123-32